\documentclass[intlimits,twoside,a4paper]{article}
\usepackage{amsmath,amssymb}
\usepackage{graphicx}

\usepackage[T2A]{fontenc}
\usepackage[cp1251]{inputenc}

\usepackage[]{cmpj2}

\issue{2016}{19}{1}{13001}
\doinumber{10.5488/CMP.19.13001}

\title[Ordering and order-disorder phase transition in the ($1\times 1$) monolayer\ldots]%
{Ordering and order-disorder phase transition in the ($1\times 1$) monolayer chemisorbed on the (111) face of an fcc crystal\thanks{Dedicated to our friend Stefan Soko{\l}owski on the occasion of his 65th birthday \protect}}
\author[A. Patrykiejew, T. Staszewski]{A. Patrykiejew, T. Staszewski}
\address{Department for the Modeling of Physico-Chemical Processes,
Maria Curie-Sk{\l}odowska University, \\
20-031 Lublin, Poland}

\date{Received September 13, 2015, in final form October 28, 2015}
\authorcopyright{A. Patrykiejew, T. Staszewski, 2016}

\begin{document}

\newcommand{\mtau}{\mbox{\boldmath$\tau$}}
\newcommand{\mqu}{\mathbf{q}}
\newcommand{\mbr}{\mathbf{r}}
\newcommand{\mbR}{\mbox{\boldmath$R$}}
\newcommand{\mbu}{\mathbf{u}}

\maketitle

\begin{abstract}
In this paper we have considered a simple lattice gas model of chemisorbed monolayer which
allows for the harmonic fluctuations of the bond length between the adsorbate atom and the surface site.
The model also involves a short-ranged attractive potential acting between the adsorbed atoms as
well as the surface periodic corrugation potential. It has been assumed that the adsorbed atoms are
bonded to the uppermost layer of the substrate atoms. In particular, using Monte Carlo simulation method
we have focused on the orderings appearing in the dense monolayer formed on the
(111) face of an fcc solid. Within the lattice gas limit, the chemisorbed layer forms a ($1\times 1$)
structure. On the other hand, when the bonds are allowed to fluctuate, three other different ordered
phases have been found to be stable in the ground state. One of them has been found to be
stable at finite temperatures and to undergo a phase transition to the disordered state.
The remaining two ordered states have been found to be stable in the ground state only. At finite
temperatures, the ordering has been demonstrated to be destroyed due to large entropic effects.
\pacs 02.50.Ng, 05.70.Fh, 68.43.Hn, 65.80-g
\end{abstract}

\section{Introduction}

The ever-growing technological importance of the on-demand tailored nanomaterials, makes it of
great importance to understand and master their formation, structure and thermodynamics as well as
electronic properties. Nanostructures of reduced geometry, like finite adsorbed islands,
have become a field
of intensive research in the last decades \cite{nano1,nano2,nano3}.
Owing to the development of powerful experimental methods, like the scanning tunneling
microscopy (STM), it is now possible to study the inner structure of even small adsorbate clusters and
to determine the arrangement of individual atoms within such nanoscopic islands \cite{stm1,stm2,stm3,stm4,stm5}.

Order-disorder phenomena in strongly adsorbed and chemisorbed monolayers have been a
subject of active research for many years now \cite{ord1,ord2,ord3,ord4,ord5,ord6,ord7,chlat1,chlat2,chlat3}
and those studies have vastly relied on Monte Carlo simulations carried out in a general framework of lattice gas
models \cite{ord1,ord2,ord7}.

At present, theoretical studies of chemisorption are usually based on {\it ab initio} quantum mechanical
calculations, and employ various methods, like the density functional theory \cite{dft1,dft2,dft3}. Although
quantum mechanical calculations can be carried out only for rather small clusters, they nonetheless allow for
a precise determination of interaction potentials and the structure of different ordered phases in chemisorbed
layers. On the other hand, ab-initio approaches cannot  be still efficiently used to study cooperative
phenomena and phase transitions in particular. Such studies require rather large systems consisting of
$10^3-10^4$ atoms and are beyond the reach of {\it ab initio} molecular dynamics \cite{md1,md2}
and Car-Parrinello \cite{car1,car2} methods. Therefore, one has to use classical simulation
methods with appropriately tuned interaction potentials \cite{clas-mc,clas-md}.

In the case of weakly adsorbed layers, made of simple gases (e.g., Ar, Kr, Xe, N$_2$, CH$_4$) on
various solid substrates (e.g., graphite, boron nitride and metal crystals), the
adsorbate-adsorbate and the adsorbate-substrate interactions can be quite well described by the
Lennard-Jones potential \cite{ads1,ads2}.  This potential does not properly describe the interactions
operating in the systems involving metals and/or semiconductors. For metal-metal interactions, the
embedded atom method (EAM) provides an approach allowing for the development of potentials with numerous
applications in theoretical and computer simulation studies of single metals and metallic
alloys \cite{eam1,eam2,eam3,eam4}. For covalently bonded systems, the potentials proposed by Stillinger and
Weber \cite{sw1,sw2}, by Tersoff \cite{Ter1,Ter2} and by Brenner \cite{Bre1} are often used.

In this work, we consider a simple lattice-like model of finite chemisorbed monolayer films
that involves the pair adsorbate-adsorbate interaction,
the harmonic potential that accounts for possible displacement of adsorbed atoms from the active
centers to which the chemisorbed atoms are bonded, and
the corrugation potential resulting from the lattice structure of the crystalline substrate \cite{ads1,cor2}.

The idea of including translational degrees of freedom (elastic interaction) into lattice gas models is not new.
Such models have been used to study the phase behavior and unmixing transition in Si-Ge
alloys \cite{lan1}, Ising ferro- and antiferromagnets \cite{lan2,lan3,lan4}, adsorbed monolayers \cite{lan5}
and frustrated materials \cite{elast-lat}.

Although the results presented in this paper should be considered as preliminary and have been obtained
for a very simple version of the model, nevertheless they show novel forms of ordering that may appear
in chemisorbed layers.

The model can be readily extended to take into account three-body interactions,
the possibility of the appearance of different types of active sites at the substrate
surface, as well as different symmetry properties of both the adsorbing surface and the
adsorbed layer. Thus, it holds promise of having predictive power
for a broad range of questions.

\section{The model}
%

We assume that the potential representing the pair interaction between adsorbate atoms
takes the following form:

\begin{equation}
u(r) = \left\{\begin{array}{ll}4\varepsilon\left[(\sigma/r)^{12} - (\sigma/r)^{6}\right] \exp\left[-\sigma/(r-R_\textrm{c})\right], & r<R_\textrm{c}, \\
                0, & r\geqslant R_\textrm{c},
         \end{array}
\right.
\label{eq:01}
\end{equation}
i.e., it is a standard Lennard-Jones (12,6) potential supplemented by the cut-off function,
with $R_\textrm{c}$ being the parameter determining the range of interaction. Equation (\ref{eq:01})
can be treated as a special case of the pair interaction term that appears in the Stillinger-Weber potential \cite{sw1}.

We also assume that all adsorbed atoms are bonded to adsorption centers, which form a
two-dimen\-sional lattice of the symmetry imposed by the substrate structure.
Throughout this work, we have considered the surface lattice of
triangular symmetry and have taken the surface lattice constant $a$ as a unit of length.
A further assumption is that $\sigma<a$ so that all lattice sites
are accessible to adsorption. In this work, we are not interested in the adsorption process but only
in the structure and properties of films in which all sites are occupied by the adsorbed atoms.
Thus, each surface site, located at
$\mbr_{0,i}$, is occupied by an adatom. Nonetheless, the bonds between adatoms and surface sites have been assumed
not to be completely rigid, but permitted to undergo small harmonic deformations. Therefore, the adatom energy
changes with the displacement from the lattice site, $\mbu= \mbr-\mbr_{0,i}$, as follows:
\begin{equation}
u_\textrm{har}(\mbu) = \frac{1}{2}f\mbu^2,
\label{eq:02}
\end{equation}
where $f$ is the force constant of the harmonic potential.

In what follows we use the reduced units. All the energy-like parameters are given in units of $\varepsilon$, e.g., the
potential energy is given as $u^{\ast} = u/\varepsilon$ and the force constant is given by $f^{\ast}= f/\varepsilon$. All the
distances are given in units of the surface lattice constant $a$. Thus,
$r^{\ast} = r/a$, $R_\textrm{c}^{\ast} = R_\textrm{c}/a$ and $\sigma^{\ast} = \sigma/a$.

Due to the lattice nature of the crystalline surface, the surface potential
experienced by adatoms exhibits periodic variations and can be expressed as follows:
\begin{equation}
v(\mbr) = v_0\sum_k\cos[\mqu_k\mbr].
\label{eq:03}
\end{equation}
In the above, $v_0$ determines the amplitude of the corrugation potential and the sum runs over the
reciprocal lattice vectors of the surface lattice ($\mqu_k$).

When $v_0=0$, the interaction energy of the pair of atoms adsorbed over
the neighboring sites depends only upon their distance $r$, and the displacements $u_i=|\mbu_i|$, ($i=1,2$)
of both interacting atoms should be the same with the displacement vectors $\mbu_1 =-\mbu_2$.

\begin{figure}[!t]
\centerline{
\includegraphics[width=0.65\textwidth]{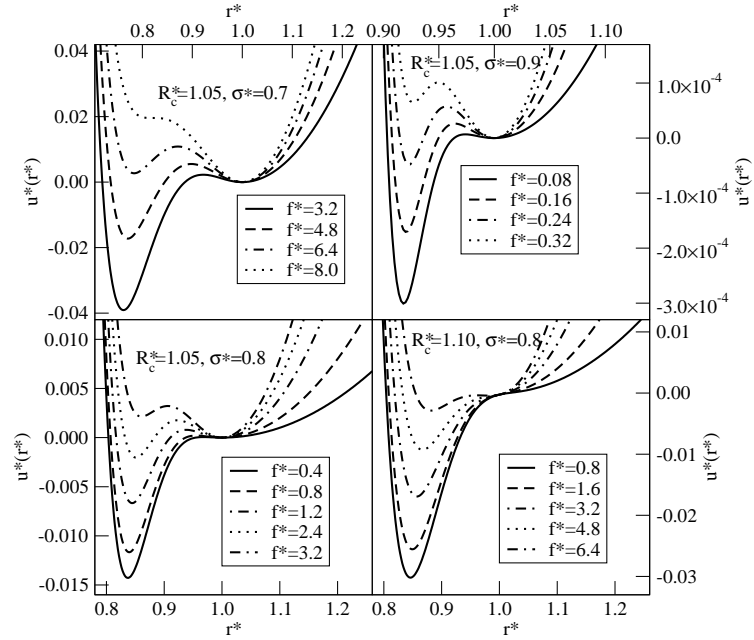}
}
\caption{The examples of the pair interaction potential $u(r)+u_\textrm{har}(|\mbu|)$ for the systems with
different $\sigma^{\ast}$, $R_\textrm{c}^{\ast}$ and the force constant $f^{\ast}$.} \label{Fig01}
\end{figure}

Figure~\ref{Fig01} gives some examples of the pair interaction potential, being the sum of $u(r)$ and $u_\textrm{har}(\mbu)$,
obtained for different sets of the parameters:
$\sigma^{\ast}=\sigma/a$, $R_\textrm{c}^{\ast}= R_\textrm{c}/a$ and $f^{\ast}=f/\varepsilon$. Due to the addition of a harmonic term,
the pair interaction energy becomes positive for large $r$ and possibly exhibits two minima.  It should
be emphasized that the pair interaction potential is
qualitatively very similar to that obtained from {\it ab initio} quantum mechanical calculations obtained
for Pb adsorbed on Si within the generalized gradient approximation (GGA) \cite{Pb-Si-1}.

Here, we assume that the adsorbed layer has been formed on the (111) face of a perfect fcc crystal and
forms a simple ($1\times 1$) structure when the bonds are strictly rigid ($f^{\ast}=\infty$).

\section{The ground state behavior}

The ground state behavior of the model depends on the assumed lattice symmetry, the pair potential cutoff distance ($R_\textrm{c}^{\ast}$),
the diameter of adatoms ($\sigma^{\ast}$), the value of the force constant $f^{\ast}$, and the amplitude of the surface
potential $v_0^{\ast}$.

\begin{figure}[!t]
\centerline{
\includegraphics[width=0.55\textwidth]{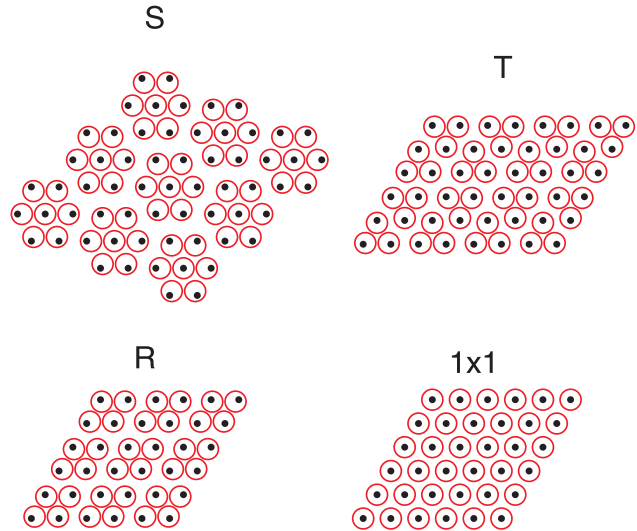}
}
\caption{(Color online) The four different ordered structures found for the systems characterized by different force constant.
Small filled circles are the locations of surface active sites while big circles represent the positions of adatoms.} \label{Fig02}
\end{figure}

\begin{figure}[!b]
\centerline{
\includegraphics[width=0.55\textwidth]{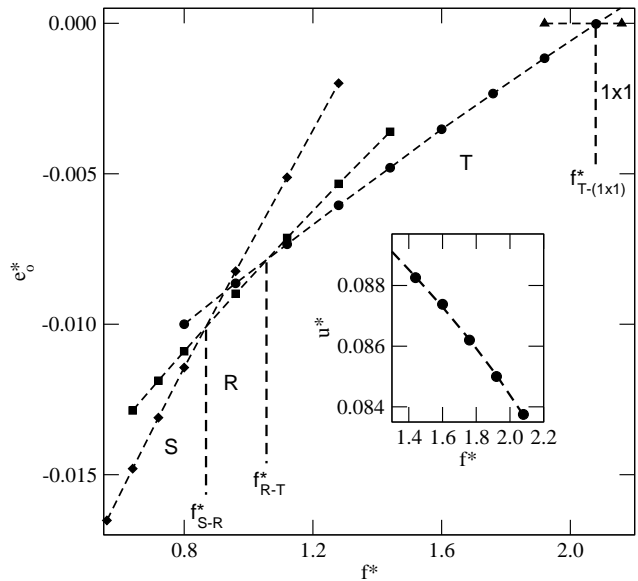}
}
\caption{The ground state phase diagram, in the plane energy ($e_0^{\ast}$)~--- force constant ($f^{\ast}$)
for the systems with $\sigma^{\ast}=0.8$ and $R_\textrm{c}^{\ast}=1.05$.
The inset shows the changes of adatom displacement with $f^{\ast}$ in the ordered structure T.} \label{Fig03}
\end{figure}

The first series of calculations aiming at the determination of stable ground state structures has been performed for
the systems without the corrugation potential ($v_0^{\ast}=0$), assuming that $\sigma^{\ast}=0.80$ and
$R_\textrm{c}^{\ast} = 1.05$, and using finite clusters of different size and shape. The calculations have been carried out over a wide range of
$f^{\ast}$ between 0.4 and 3.0. Four different
ordered structures, depicted in figure~\ref{Fig02}, have been found. The stability of each of them is determined by the
magnitude of the force constant  $f^{\ast}$. Figure~\ref{Fig03} shows the ground state phase diagram obtained
for the systems with $\sigma^{\ast}=0.80$ and $R_\textrm{c}^{\ast} = 1.05$ and different values of $f^{\ast}$.

In order to determine the structure of the system in the ground state, we have calculated the energy for all possible structures as
a function of the displacement. Of course, the structure $(1\times 1)$ corresponds to zero displacement. The energy of the structures S and T is
characterized by a single displacement, while the energy of the structure R is determined by two displacements in two orthogonal directions along
the diagonals of the rhombus. The stable state of a system is the one in which the energy reaches its minimum. We have determined
the displacements that minimize the energy for each structure, as well as determined which of the structures is stable for a given
set of parameters.

For sufficiently high values of $f^{\ast}$ ($f^{\ast}>f^{\ast}_{\textrm{T}-(1\times 1)})$, a simple $(1\times 1)$ structure is
a stable state. In this case, the adatoms are not displaced from the positions of adsorption centers,
i.e., the average displacement $\mbu$ is equal to zero. When $f^{\ast}$
decreases, the adatoms exhibit a gradually increasing tendency to be displaced from the positions given by
the vectors $\mbr_{0,i}$ and this leads to the formation of the structures T, R and S, depicted in
figure~\ref{Fig02}. The structure
T consists of equilateral triangles of adatoms arranged in such a way that each adatom has a displacement from the surface lattice site of the same length. Of course, the magnitude of
this displacement gradually
increases when $f^{\ast}$ becomes lower (see the inset to figure~\ref{Fig03}). The structure T is stable as long as
$f^{\ast}$ stays between $f^{\ast}_{\textrm{T}-(1\times 1)}$ and $f^{\ast}_\textrm{R--T}$. The structure R develops for the lower values of $f^{\ast}$,
between $f^{\ast}_\textrm{S--R}$ and $f^{\ast}_\textrm{R--T}$. In this structure, the groups
of four adatoms
form rhombic clusters, characterized by two different displacements of adatoms from the surface lattice
(along the diagonals). In perfectly ordered
T and R states, the unit cell vectors are aligned with the symmetry axes of the triangular lattice.
When $f^{\ast}$ falls below $f^{\ast}_\textrm{S--R}$, the structure labeled as S (star-like) becomes stable. This structure
consists of clusters made of seven atoms each and arranged in such a way that the central atom is not displaced
from the adsorption site, while its six nearest neighbors are equally displaced towards the central atom.
The central atoms form a $\sqrt{7}\times\sqrt{7}$ lattice rotated by $19.1^\circ$ with respect to the original triangular
lattice of adsorption sites. Of course, in the R and S
structures, the atomic displacements also gradually increase when the force constant $f^{\ast}$ becomes lower, just the
same as in the case of T structure. For the values of $f^{\ast}$ well below $f^{\ast}_\textrm{S--R}$, the tendency of adatoms to
be displaced from lattice sites becomes high enough to lead to the formation of structures consisting
of clusters larger than in the S structure
and this point will be addressed later on in section~5. Of course, when $f^{\ast}$ approaches zero, the
entire adsorbed island forms a close packed triangular lattice, but in the case of chemisorption, this situation
can be excluded from consideration.

\begin{figure}[!h]
\centerline{
\includegraphics[width=0.55\textwidth]{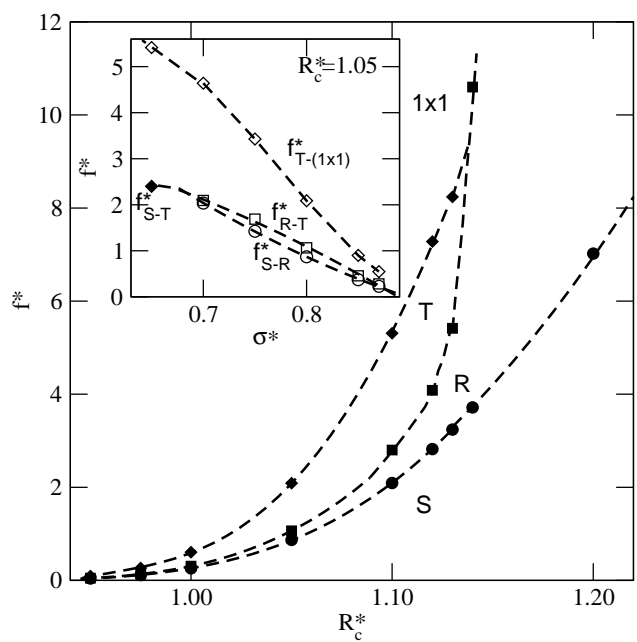}
}
\caption{The changes of the regions of stability of different ordered phases with the cut-off distance for
the systems with $\sigma^{\ast}=0.8$.} \label{Fig04}
\end{figure}

It should be noted that although our ground state calculations have been carried out for finite systems,
the results have not been affected by finite size effects. The assumed range of the interaction potential
[equation (\ref{eq:01})] is very short, and
due to rather large displacements of adatoms from lattice sites, the system energy in the ground state is
not influenced by the presence of free boundaries.
 For the potentials with a longer interaction range,
given by $R^{\ast}_\textrm{c}$, finite size effects become of importance. In such systems, the stability regions of
different ordered states are also affected by finite size and boundary effects.
The change of the interaction
range affects the strength of the pair potential, cf. the
lower panels of figure~\ref{Fig01}, and hence the values of $f^{\ast}$ that delimit the regions of stability of different
ordered states. This is illustrated by the results given in figure~\ref{Fig04}. The results show that the
T structure exists
only for rather short-ranged potentials with $R_\textrm{c}^{\ast}$ lower than about 1.139, while the R and S structures are
stable even for the potentials of a longer range. However, with the increase of $R_\textrm{c}^{\ast}$, the results are subjected
to gradually increasing finite size effects. The ground state calculations for the clusters
of perfectly ordered S structure and of different size have shown that for $f^{\ast}\leqslant 2.4$, the S structure
occurs only in finite clusters. Upon the increase of the cluster size, the displacement of adatoms corresponding to
the minimum of potential energy ($u_\textrm{min}$) goes to zero. The ground state properties of the model also depend
upon the value of $\sigma^{\ast}$. Of course, for $\sigma^{\ast}$, at which the minimum of the potential occurs at the distance
approaching $a$, only the ($1\times 1)$ structure is stable. Upon the decrease of $\sigma^{\ast}$, the location
of the potential minimum shifts towards shorter distances between the interacting atoms (cf. figure~\ref{Fig01}) allowing for rather
large atomic displacements.
This gives rise to the appearance of S, R and T structures. We have performed ground state calculations for a fixed
value of $R^{\ast}_\textrm{c} = 1.05$ and for different $\sigma^{\ast}$ between 0.65 and 0.87. The results of the calculations are
summarized in the inset to figure~\ref{Fig04}. It appears that the R structure occurs only for $\sigma^{\ast}$ larger than about
0.68. For still smaller atoms, only the S, T and ($1\times 1$) structures are stable in the ground state. Due to the lowering of
$\sigma^{\ast}$, the energy of the pair interaction increases so that for a given $f^{\ast}$ the adatoms
are more likely to exhibit larger
displacements from the lattice sites. This explains why the values of $f^{\ast}_\textrm{S--R}$, $f^{\ast}_\textrm{R--T}$ and
$f^{\ast}_{\textrm{T}-(1\times 1)}$ increase when $\sigma^{\ast}$ becomes lower.

In the case of an (111) fcc surface considered here, the effects of the corrugation potential on the ground state properties
can be readily anticipated. The adsorption centers are located right above the surface atoms, i.e., over
the maxima of the external field, and the atomic displacements in all S, R and T structures are
directed towards the positions of a lower surface potential $v(\mbr)$. Therefore, an increase of the corrugation potential
amplitude ($v_0^{\ast}$) has the same effect as the lowering of the force constant ($f^{\ast}$).
In fact, the calculations performed have demonstrated that the values of $f^{\ast}_\textrm{S--R}$,
$f^{\ast}_\textrm{R--T}$ and  $f^{\ast}_{\textrm{R}-(1\times 1)}$ linearly increase with $v_0^{\ast}$ as follows:
\begin{equation}
f^{\ast}_{\alpha-\beta}(v_0^{\ast}) = f^{\ast}_{\alpha-\beta}(v_0^{\ast}=0)+ C v_0^{\ast}
\label{eq:04}
\end{equation}
with $\alpha-\beta$ being S--R, R--T or $\textrm{T}-(1\times 1)$, and the magnitude of the constant $C$ depending
only upon $\sigma^{\ast}$ and $R_\textrm{c}^{\ast}$. In other words, due to the turning on of the corrugation potential
at zero temperature ($T=0$), the limits of stability of all differently ordered structures uniformly shift towards higher values of  $f^{\ast}$.

\section{Monte Carlo simulation method}

The above presented model has been studied using a standard Metropolis Monte Carlo simulation
method in the canonical ensemble \cite{LB2000}. We have monitored the system energy and the contributions
to the energy due to the pair interaction represented by the potential [$u(r)$], the harmonic
interaction [$u_\textrm{har}(\mbu)$] and due to the corrugation potential [$v(\mbr)$]. Furthermore, we have
calculated the heat capacity from the energy fluctuations.

In order to monitor the formation of the ordered structures, we have calculated the probability distribution
of atomic displacements $p_\textrm{d}(|\mbu|)$ and the probability distribution of the number of nearest
neighbors $p_{NN}(n)$. Since
all surface lattice sites are covered by adatoms, we have
assumed that the nearest neighbors are only those atoms located on the nearest adsorption sites with
the inter-atomic distance smaller than unity. Thus, only the atoms displaced from lattice sites have
been counted.

In the case of the ordered T structure, the distribution $p_\textrm{d}(|\mbu|)$ should be unimodal and $p_{NN}(n)$
should be close to unity for $n=2$, while approaching zero for other values of $n$. The structure
S should give bimodal distribution $p_\textrm{d}(|\mbu|)$; with one maximum at $|\mbu|\approx 0$ (for the central atoms)
and the second maximum at larger value of $|\mbu|$ (for the nearest neighbors of the central atom). The
distribution $p_{NN}(n)$ should be equal to about 1/7 for $n=6$, about 6/7 for $n=3$ and approach zero for
other values of $n$. Finally,
the structure R should also be characterized by bimodal distribution $p_\textrm{d}(|\mbu|)$ with the maxima at the
positions corresponding to the two displacements in the rhombus, while the distribution $p_{NN}(n)$
should approach the value of 0.5 for $n=2$ and 3 and should be approximately zero for other values of $n$.

The calculations have been performed using two different shapes of the simulation box. One series of
calculations has been carried out using simple rhombic cells
oriented along the symmetry axes of the triangular lattice and of different side length, $L$ (labeled as R-cells).
Note
that the ordered structures T and R require even values of $L$ to accommodate them within the cell. In the
case of the ordered structure S, we have also used rhombic simulation cells, but appropriately rotated and
of the side length adjusted to accommodate the S structure (labeled as S-cells).

\section{Results and discussion}

The primary goal here has been to study the mechanism of disordering of the structures S, R and T.
Under the assumption of strong bonding between adatoms and surface sites, the desorption
has been prohibited so that the disordering can only involve the destruction of the order specific
to S, R and T structures.
We have performed Monte Carlo simulations for several systems using finite R- and S-cells of different
size. Taking into
account that one of our goals was to study truly nanoscopic systems, we have not applied periodic boundary
conditions.

\begin{figure}[!h]
\centerline{
\includegraphics[width=0.65\textwidth]{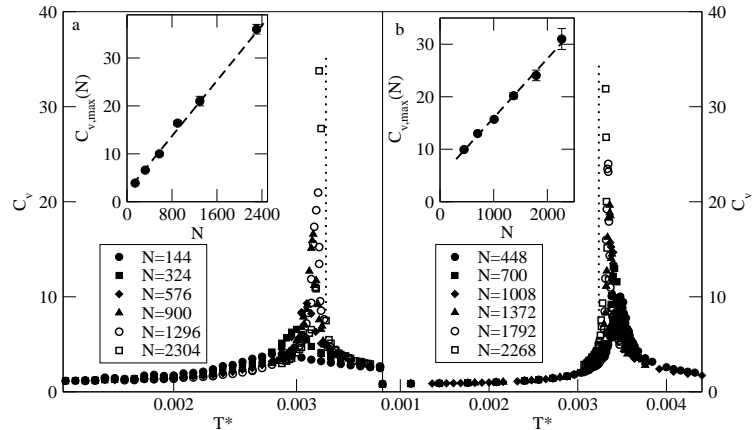}
}
\caption{Main panels show the heat capacity curves for the system with $\sigma^{\ast}=0.8$,
$R_\textrm{c}^{\ast} =1.05$ and $f^{\ast}=0.4$ for different sizes of the adsorbed island (given in the figure).
The left-hand  panel (a) gives the results for the simple rhombic simulation cell and the right-hand panel (b) corresponds
to the simulation cell accommodating the S structure. The insets show the scaling plots of the  heat capacity
maximum versus system size.} \label{Fig05}
\end{figure}

\begin{figure}[!t]
\centerline{
\includegraphics[width=0.6\textwidth]{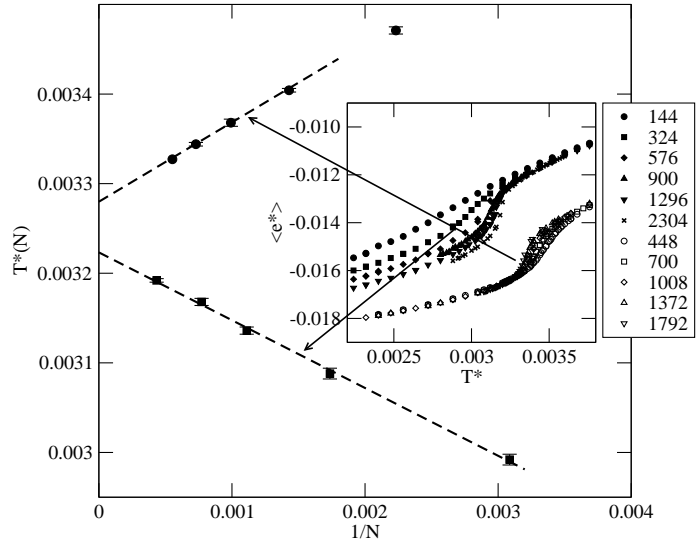}
}
\caption{The changes of the temperature at which the heat capacity reaches its maximum plotted against the number of atoms
in the adsorbed island for the systems  with $\sigma^{\ast}=0.8$, $R_\textrm{c}^{\ast} =1.05$ and $f^{\ast}=0.4$.
Filled circles correspond to the simulation cell perfectly accommodating the S structure, while the filled squares, to the
simple rhombic box. The inset shows the changes of the average potential energy $\langle e^{\ast}\rangle$ with temperature $T$ for both cases and
different size of the adsorbed island.} \label{Fig06}
\end{figure}

At first, we have considered the systems with $\sigma^{\ast}=0.8$ and $R_\textrm{c}=1.05$ and the values of $f^{\ast}$
for which the ordered S structure is stable in the ground state, i.e., with $f^{\ast}<f^{\ast}_\textrm{S--R}=0.864$.
 Parts (a) and (b) of figure~\ref{Fig05} show the heat
capacity curves obtained for the system with $f^{\ast}=0.4$. Part (a) of figure~\ref{Fig05} gives the results obtained using
the R-cells of different size, while part (b) gives the results obtained for finite S-cells,
also of different size.
Both series of calculations have demonstrated that the heat capacity exhibits sharp maxima
of the height and location depending upon the system size, indicating the presence of a
phase transition. The insets to parts (a) and (b) clearly
show that the maximum value of the heat capacity changes linearly  with the
number of atoms in the system $N$, i.e., with the system volume, since all lattice sites are occupied by the adsorbate
atoms. According to the
finite-size scaling theory of phase transitions \cite{LB2000,fin1}, this
sort of behavior is characteristic of the first order transition. The theory also predicts the
transition temperature in finite systems $T_\textrm{tr}(N)$ to be shifted with respect to the
transition temperature in the infinite system, $T_\textrm{tr}(\infty)$, and the following relation should hold:
\begin{equation}
T_\textrm{tr}(N) = T_\textrm{tr}(\infty) + W/N,
\label{eq:scala1}
\end{equation}
where the constant $W$ is proportional to $1/(E_+-E_-)$ with $E_-$ and
$E_+$ being the internal energies of the coexisting phases at $T\rightarrow T_\textrm{tr}^-$
and $T\rightarrow T_\textrm{tr}^+$, respectively \cite{fin2}. The difference $E_+-E_-$ is positive and hence the
temperature $T^{\ast}_\textrm{tr}(N)$ should increase with $1/N$. This does occur when the S-cells accommodating
the ordered
S structure are used (see figure~\ref{Fig06}). On the other hand, the
temperature $T^{\ast}_\textrm{tr}(N)$ decreases with $1/N$ when the R-cells are used. In this case,
the ordering is not perfect. The regions close to the island boundaries do not
show the ordering characteristic of the S structure, even at very low temperatures. At higher temperatures,
the disorder easily propagates into the
system interior and considerably affects the estimated transition temperature. Of course, these effects of
propagated disorder gradually decrease when the simulation cell size becomes larger, so that the transition temperature
gradually increases with the system size. This is very well illustrated by the curves showing the changes of the system
energy with temperature and obtained for different system sizes (see the inset to figure~\ref{Fig06}). In the case of simulation
with S-cells, the energy at low and high temperatures does not depend upon the system size at all.
Only at the temperatures close to the phase transition,
the size effects set in. In fact, the results obtained from the runs with
and without periodic
boundary conditions applied have given the same results. On the other hand, the energy curves obtained for R-cells
demonstrate large finite size effects already at low temperatures. This is a consequence of the already mentioned
disordering close to the island boundaries.  Therefore, only the simulation
with S-cells gives the correct results and the estimated transition temperature $T_\textrm{tr}(\infty)$
is equal to about 0.00328.

\begin{figure}[!b]
\centerline{
\includegraphics[width=0.6\textwidth]{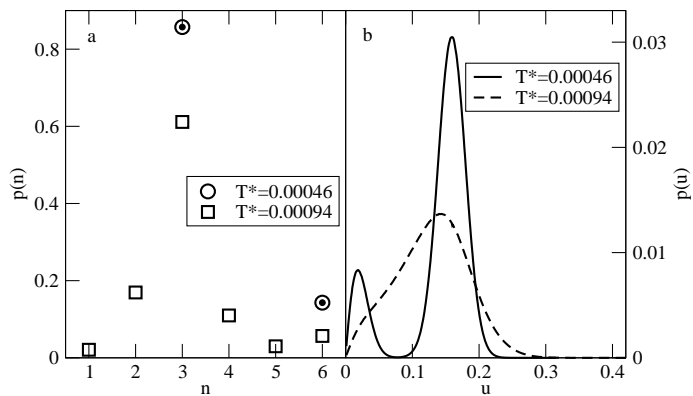}
}
\caption{The probability distribution functions $p(n)$ and $p(|\mbu|)$ for the system with $\sigma^{\ast}=0.8$,
$R_\textrm{c}^{\ast} =1.05$, and $f^{\ast}=0.4$ at the temperature below and above the order-disorder transition point
obtained for the adsorbed island containing 1372 atoms. The solid circles in the left-hand panel
show the probabilities of an atom to have 3 and 6 nearest neighbors in the  perfect S structure.} \label{Fig07}
\end{figure}

The nature of the transition becomes quite clear when one considers the behavior of the probability distributions
$p_\textrm{d}(|\mbu|)$ and
$p_{NN}(n)$ given in figure~\ref{Fig07}.
The distributions recorded below the transition temperature show a perfect ordering into the S structure, while
those obtained at the temperatures above the transition point indicate that the disordering leads to the destruction
of the S structure. In particular, the distribution
$p_\textrm{d}(|\mbu|)$ looses its bimodal shape and the distribution $p_{NN}(n)$ shows quite large contributions
due to different numbers of nearest-neighbors, other than 3 or 6.

\begin{figure}[!t]
\centerline{
\includegraphics[width=0.52\textwidth]{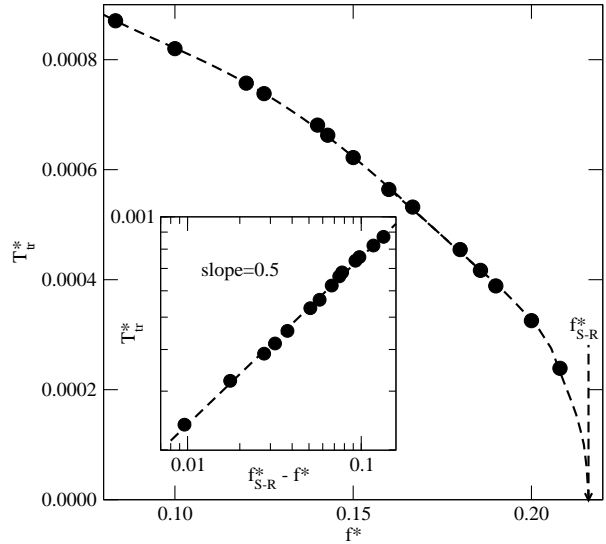}
}
\caption{The main figure shows the changes of the transition temperature for the S structure versus $f^{\ast}$
obtained for the systems with $\sigma^{\ast}=0.8$ and $R_\textrm{c}^{\ast} =1.05$. The inset
shows the scaling plot of the transition temperature versus $f^{\ast}_\textrm{S--R}-f^{\ast}$.} \label{Fig08}
\end{figure}

We have carried out simulations for a series of systems with different values of the force constant, $f^{\ast}$,
and estimated the transition temperatures. The results of our calculations have been summarized in figure~\ref{Fig08}, which
illustrates the changes of $T^{\ast}(\infty)$ with $f^{\ast}$. As expected, the transition temperature ($T^{\ast}_\textrm{tr}$)
gradually decreases towards zero when $f^{\ast}$ grows towards $f^{\ast}_\textrm{S--R}$,

\begin{figure}[!b]
\centerline{
\includegraphics[width=0.51\textwidth]{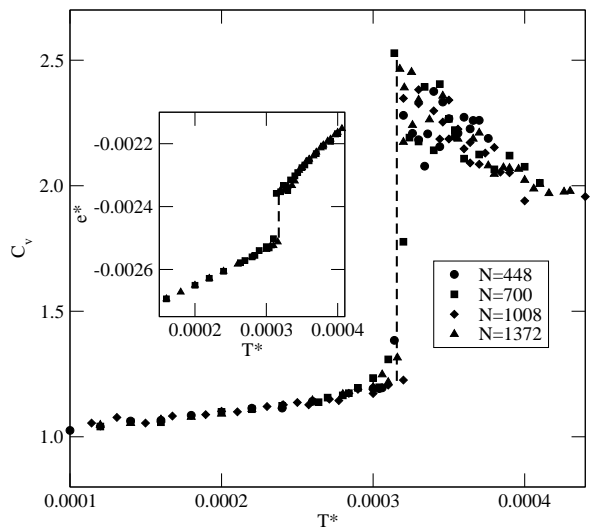}
}
\caption{The main figure shows the heat capacity curves for the system with $\sigma^{\ast}=0.8$,
$R_\textrm{c}^{\ast} =1.05$ and $f^{\ast}=0.8$ and different size of the ordered into the S structure adsorbed islands. The inset shows
the changes of the potential energy versus temperature for the same system. The vertical dashed line marks the
location of the order-disorder transition.} \label{Fig09}
\end{figure}

The log--log plot of $T^{\ast}_\textrm{tr}$ versus $\Delta f^{\ast}= f^{\ast}_\textrm{S--R} - f^{\ast}$
(given in the inset to figure~\ref{Fig08}) demonstrates
that there are three regions of $\Delta f^{\ast}$ over which the system exhibits different behavior.
For small values of $\Delta f^{\ast}$, i.e., for the values of $f^{\ast}$ close to $f^{\ast}_\textrm{S--R}$,
the first-order character of the transition
becomes more pronounced, for example when $f^{\ast}=0.72$, and 0.8 the behavior of
the heat capacity and of the potential energy is different than in the previously discussed case of $f^{\ast}=0.4$, as
well as for other values of $f^{\ast}$ up to 0.64. Both quantities exhibit
discontinuities at the transition temperature (see figure~\ref{Fig09}) without showing any systematic finite size effects.

For intermediate values of $\Delta f^{\ast}$ between about 0.64 and 1.6, the transition is only weakly first-order
and exhibits the behavior already described while discussing the system with $f^{\ast}=0.4$.

\begin{figure}[!t]
\centerline{
\includegraphics[width=0.5\textwidth]{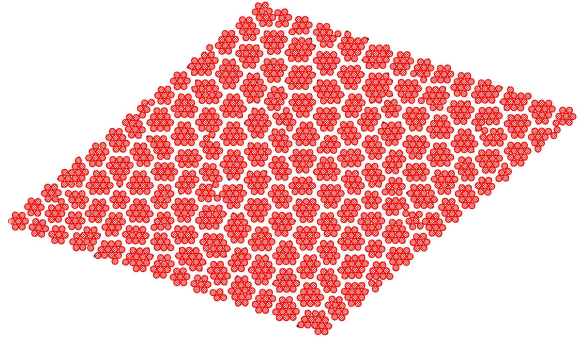}
}
\caption{(Color online) A snapshot of the configuration obtained for the system with
$\sigma^{\ast}=0.8$, $R_\textrm{c}^{\ast} =1.05$ and $f^{\ast}=0.16$ at $T^{\ast}=0.0024$.} \label{Fig10}
\end{figure}

\begin{figure}[!b]
\centerline{
\includegraphics[width=0.5\textwidth]{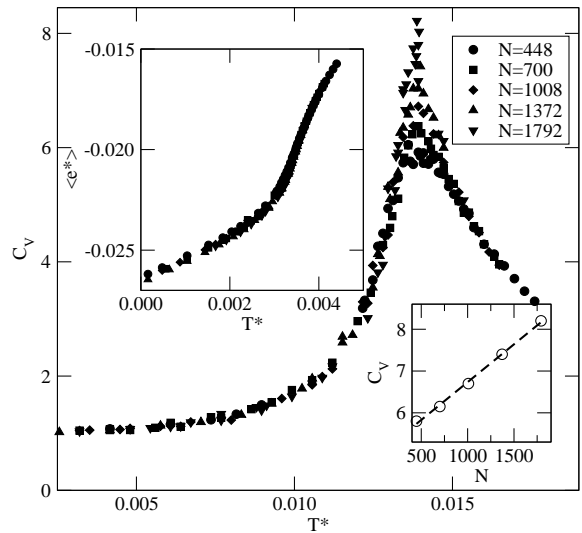}
}
\caption{The temperature changes of the heat capacity (main figure) and of the potential energy
(the upper inset) for the system with $\sigma^{\ast}=0.8$, $R_\textrm{c}^{\ast} =1.05$ and $f^{\ast}=0.16$. The lower inset
shows the changes of the heat capacity maximum with the system size.} \label{Fig11}
\end{figure}

For sufficiently large $\Delta f^{\ast}$, i.e., for sufficiently small $f^{\ast}$, the structure of adsorbed layer
changes, since adatoms enjoy much more freedom to displace from the surface lattice
sites. Consequently, the attractive pair potential wins over the harmonic forces and
enhances the tendency towards clustering. Hence, adatoms group into lager clusters than those in the S
structure. This is well illustrated by the snapshot given in figure~\ref{Fig10}, recorded at $T^{\ast}=0.0024$
for the system with $f^{\ast}= 0.16$ and consisting of
1792 atoms. One can see that these clusters may contain a variable number of atoms and
be of different shape. We have studied this system using R- and S-cells with periodic boundary
conditions applied. The heat capacity and potential energy curves obtained for systems of different size have
demonstrated (see figure~\ref{Fig11}) that the disordering occurs over a rather narrow temperature range, though the
transition region is considerably smeared and the heat capacities obtained for systems of rather large size
do not reach such high values as observed in the systems ordering into the S structure. Nonetheless, the heat
capacity maxima increases linearly
 with $N$, suggesting that the system undergoes a weakly first-order phase transition. The transition
occurs between the structure consisting of a large number of small clusters and the disordered state
with more-or-less uniform distribution of atoms.

\begin{figure}[!t]
\centerline{
\includegraphics[width=0.65\textwidth]{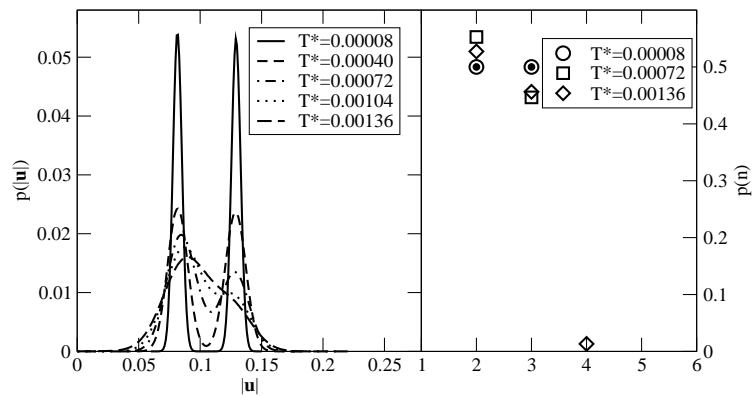}
}
\caption{The probability distribution functions $p(n)$ and $p(|\mbu|)$ for the system with $\sigma^{\ast}=0.8$,
$R_\textrm{c}^{\ast} =1.05$ and $f^{\ast}=0.96$ at different temperatures obtained for the adsorbed island containing 1372 atoms.
The solid circles in the left-hand panel show the probabilities of an atom to have 2 and 3 nearest neighbors in
the perfect R structure.} \label{Fig12}
\end{figure}

As soon as the value of the force constant becomes higher than $f^{\ast}_\textrm{S--R}$, the systems are expected
to order into the R structure at low temperatures. The simulation has demonstrated that this does occur.
Figure~\ref{Fig12} presents the
examples of distribution functions $p_\textrm{d}(|\mbu|)$ and $p_{NN}(n)$ obtained at different temperatures for the
system with $f^{\ast}=0.96$. It is well seen that only at very low temperatures does the ordering into the R structure
occur. Upon an increase of temperature, the bimodal distribution $p_\textrm{d}(|\mbu|)$ gets distorted and then disappears at all. This
is accompanied by the changes in the distribution $p_{NN}(n)$. In particular, the probability for an atom to have
two nearest neighbors increases while the probability of having three nearest neighbors decreases. This
indicates that triangular clusters made of three atoms appear at elevated temperatures.
This has been confirmed by the inspection of snapshots, which demonstrated the formation of
isolated triangles, as well as a loss of alignment of rhombic clusters. In particular, the triangles have been
observed to appear close to the regions containing differently oriented rhombic clusters.
Upon an increase of temperature, the degree of disordering also gradually increases.
The heat capacity and potential energy curves (not shown here)
have clearly demonstrated that
the R structure disorders continuously and does not undergo any order-disorder phase transition. In particular, neither
the potential energy nor the heat capacity show any finite size effects.
In a perfectly ordered R phase, all rhombic clusters made of four atoms each are aligned as shown in figure~\ref{Fig02}.  Our simulation has
shown that this alignment is destroyed as soon as the temperature is raised above 0. The energy cost associated with
changing the orientations of rhombic clusters and the formation of isolated triangles is very low, while
the entropic effects are very large and destroy the alignment completely.

\begin{figure}[!t]
\centerline{
\includegraphics[width=0.65\textwidth]{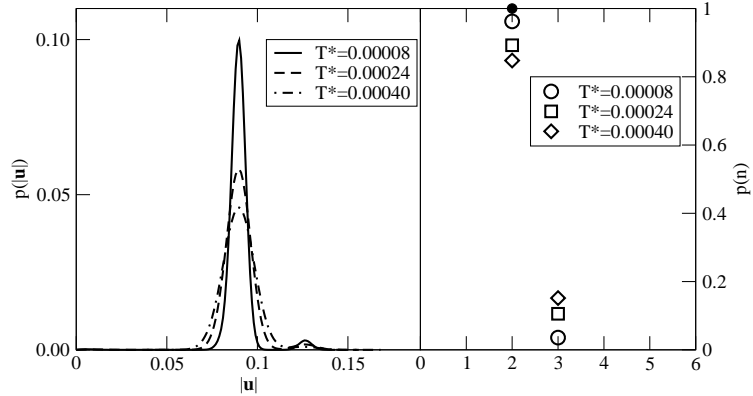}
}
\caption{The probability distribution functions $p(n)$ and $p(|\mbu|)$ for the system with $\sigma^{\ast}=0.8$,
$R_\textrm{c}^{\ast} =1.05$ and $f^{\ast}=1.12$ at different temperatures obtained for the adsorbed island containing 1372 atoms.
The solid circles in the left-hand panel show the probabilities of an atom to have 2 nearest neighbors in
the perfect T structure.} \label{Fig13}
\end{figure}

Qualitatively similar results have been obtained for the systems characterized by the values of $f^{\ast}$ higher than
$f^{\ast}_\textrm{R--T}$, i.e., when the ordering into the T structure occurs in the ground state.
Similarly to the R structure,
the disordering of the T structure gradually occurs and is accompanied by the loss of alignment of
triangular clusters and by the appearance of rhombic clusters at nonzero temperatures. In fact, already the
simulation performed at a very low temperature of $T^{\ast}=0.00008$ and
using the starting configuration being a perfect T structure has demonstrated a certain degree of disordering.
Figure~\ref{Fig13} shows the
probability distributions $p_\textrm{d}(|\mbu|)$ and $p_{NN}(n)$ recorded at very low temperatures that
demonstrate the above mentioned  partial disordering. In particular, the distribution $p_\textrm{d}(|\mbu|)$
exhibits a small second maximum at large displacements and the distribution $p_{NN}(n)$ shows a
non-zero probability of an atom to have three nearest neighbors already at $T^{\ast}=0.00008$.

We have also performed calculations for the systems with higher potential cutoff, with $R_\textrm{c}^{\ast}=1.14$, as well
as for the systems with different size of adsorbed atoms, with $\sigma^{\ast}=0.85$ and 0.7, but apart
from the obvious quantitative differences resulting from the changes in the pair potential properties, the
results have been qualitatively the same as for the above discussed systems.

\begin{figure}[!b]
\centerline{
\includegraphics[width=0.55\textwidth]{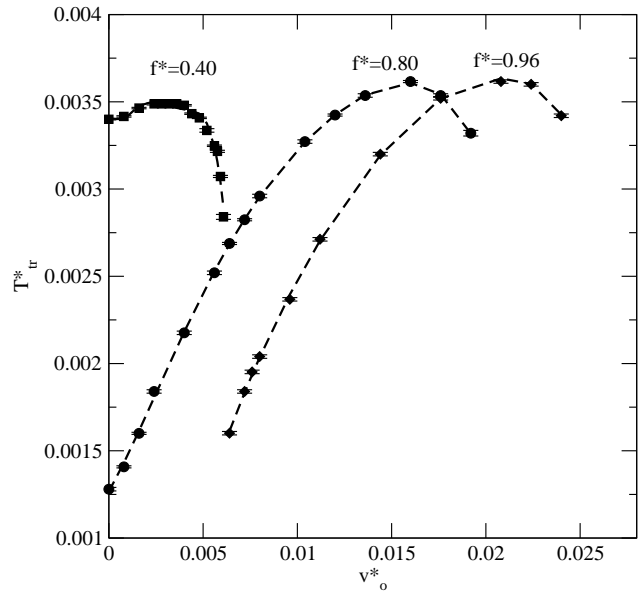}
}
\caption{The order-disorder transition temperature of the S phase versus the amplitude of the corrugation
potential obtained for the systems with $R^{\ast}_\textrm{c}=1.05$, $\sigma=0.8$ and different values of $f^{\ast}$ (given
in the figure).} \label{Fig14}
\end{figure}	

As already discussed in section~3, the corrugation potential is not expected to lead to important qualitative
changes in the behavior of the systems considered here. Indeed, our calculations carried out for several
systems have shown that the increase of $v_0^{\ast}$ is somehow equivalent to the decrease of $f^{\ast}$. It has been
already shown that for $v_0^{\ast} = 0.0$ and the values of $f^{\ast}$ corresponding to the stability
region of the S structure, a gradual decrease of $f^{\ast}$ leads to an increase of the order-disorder transition
temperature (cf. figure~\ref{Fig08}) and the stability of the $S$ structure lowers only for sufficiently
 low values of $f^{\ast}$.
 Thus, one expects that for a fixed value of $f^{\ast}$, an increase of the corrugation
potential amplitude ($v_0^{\ast}$) should also lead to a gradual increase of the transition temperature.
The calculations carried out for finite clusters  consisting of 700 atoms  with $\sigma^{\ast}=0.8$,
$R_\textrm{c}^{\ast} = 1.05$, different values of $f^{\ast}$ between $0.4$ and $0.96$ and for several values of $v_0^{\ast}$
have demonstrated a gradual increase of the transition temperature over a certain range of $v_0$ (see figure~\ref{Fig14}).
However, when the amplitude $v_0^{\ast}$ becomes high enough, the transition temperature decreases rather sharply.
 This behavior is quite similar to the one observed for the systems without the corrugation potential
and sufficiently low values of $F^{\ast}$.

It is interesting to note that in the case of $f^{\ast}=0.96$, for which a uniform system ($v_0^{\ast}=0)$
orders into the R structure in the ground state, the S structure appears to be stable when $v_0^{\ast}$ exceeds
a certain threshold value (marked by a vertical dashed line in figure~\ref{Fig14}). This clearly shows that an increase of
$v_0^{\ast}$ is equivalent to the decrease of $f^{\ast}$.

\begin{figure}[!t]
\centerline{
\includegraphics[width=0.45\textwidth]{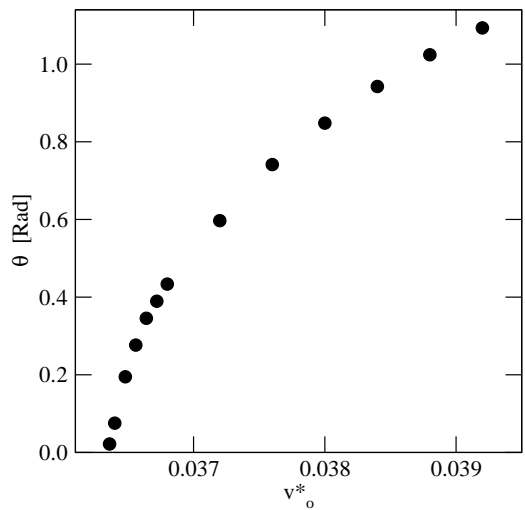}
}
\caption{The rotation angle of a single 7-atom cluster versus the amplitude of the surface potential in
the ground state of the system with  $R^{\ast}_\textrm{c}=1.05$, $\sigma=0.8$ and $f^{\ast}=0.4$.} \label{Fig15}
\end{figure}

Taking into account the properties of the corrugation potential generated over the (111) face of an
fcc crystal, one expects the 7-atom clusters constituting the S structure to be slightly rotated around
the central atom. This rotation should result from the fact that
the displacement vectors in the uniform case are not directed towards the surface potential minima,
but rather towards the saddle points. The rotation of the entire cluster
does not cost any energy of the pair interaction, but affects the harmonic and surface interactions.
The harmonic interaction hinders the cluster rotation, so that nonzero rotational angles may
occur only when the energy gain due to the corrugation potential is high enough.
The ground state calculations for the systems characterized by rather low $f^{\ast}=0.4$ have shown,
however, that even for the amplitude of the surface potential well above the upper limit of the
S structure stability and equal to about 0.0062, the cluster rotation does not occur. The rotation of a single
7-atom cluster has been found to be present only when $v_0^{\ast}$ exceeds the value of about 0.0364
(see figure~\ref{Fig15}). In the case of large systems consisting of many clusters, the clusters consisting of more
than 7 atoms begin to develop already for
$v_0^{\ast}$ larger than about 0.0062.

\section{Final remarks}

In this paper, we have discussed the ordering appearing in dense chemisorbed
monolayer films formed on the (111) fcc lattice. It has been demonstrated that when the
pair potential is short ranged and the bonds between the chemisorbed atoms and the surface atoms
are not rigid, the film may form different ordered phases. The ordering strongly
depends upon the elasticity of the bonds. When the bonds are very stiff, the film
orders into a simple $(1\times 1)$ phase. Upon a decrease of the bond stiffness,
the adatoms exhibit a gradually increasing tendency to be displaced from surface sites.
Since the pair potential is attractive at small distances, the adatoms tend to form
small clusters. In the ground state, we have found the ordered states in which
such clusters consist of three (T), four (R) and seven (S) atoms (R) (cf. figure~\ref{Fig02}).

The structures T and R have been demonstrated to be stable only
in the ground state. At finite temperatures, large entropic effects destroy the orderings.
On the other hand, the structure S has been found to be stable at finite temperatures.
Our Monte Carlo simulation results, supported by finite-size scaling analysis,
have shown that the S phase is disordered via the first-order transition. The transition
temperature depends on the bond elasticity.

For sufficiently low values of the harmonic potential force constant, the adatoms enjoy
a rather large freedom to displace from adsorption sites and tend to form
still larger clusters.

Although the model considered here is very simple, nevertheless it exhibits interesting physics,
novel phase behavior and new types of ordering. In this work, we have considered only the case of
a fully filled triangular lattice, but the model can be also used to study the orderings on the surfaces of
different symmetry of adsorption sites. It can be also readily extended to take into account three body
forces \cite{sw1} and orientation-dependent interactions \cite{sw3}.

Monte Carlo methods can be also used to
study the adsorption processes and possible different orderings appearing for lower coverages.
This requires simulations in the grand canonical ensemble, commonly used in the studies of adsorption
phenomena \cite{ord6}.

\ukrainianpart

\title
{Впорядкування і фазовий перехід порядок-безлад у ($1\times 1$) моношарі, хемосорбованому на грані (111)  fcc кристалу}
\author{А. Патрикєєв, Т. Сташевський}
\address{Відділ моделювання фізико-хімічних процесів, Університет Марії Кюрі-Склодовської, Люблін, Польща
}
\makeukrtitle

\begin{abstract}
В цій статті ми розглянули просту модель граткового газу хемосорбованого моношару,  яка дозволяє здійс\-нювати гармонічні флуктуації довжини
 зв'язку між атомами адсорбату та поверхні. Дана модель містить короткосяжний потенціал притягання, який діє між адсорбованими
 атомами, а також поверхневий періодичний гофрований потенціал. Зроблено припущення, що адсорбовані атоми є зв'язані з найвищим
 шаром атомів підкладки. Зокрема, використовуючи метод моделювання Монте Карло, акцент зроблено на впорядкувані у щільному моношарі, сформованому на
(111) грані  fcc кристалу. В границі граткового газу хемосорбований шар утворює ($1\times 1$) структуру.
З іншого боку, коли наявні флуктуації зв'язків, три інші впорядковані фази є стійкими в основному стані.
Встановлено, що одна з них є стійкою при скінчених температурах і переходить у невпорядкований стан.
  Дві інші впорядковані фази є стійкими лише в основному стані. Показано, що при скінчених температурах впорядкування руйнується завдяки сильним ентропійним ефектам.
\keywords хемосорбція, фазові переходи, комп'ютерне моделювання, наноскопічні системи
\end{abstract}

\end{document}